\documentclass[a4paper,12pt]{article}
\usepackage{fullpage}
\usepackage[T1]{fontenc}
\usepackage[utf8]{inputenc}
\usepackage{lmodern}
\usepackage[british,english]{babel}
\usepackage{multicol}
\usepackage{setspace}
\usepackage[parfill]{parskip}
\let\OLDthebibliography\thebibliography
\renewcommand\thebibliography[1]{
  \OLDthebibliography{#1}
  \setlength{\parskip}{0pt}
  \setlength{\itemsep}{0pt plus 0.3ex}
}
\title{Freedom From Restriction, Freedom Of A Restriction: A Comparison Of Some Open Source Software Licenses}
\author{Martin A.~COLEMAN\footnote{Copyright \copyright\ 2014 Martin A. COLEMAN. This document may be freely shared, copied, transferred and/or re-distributed, in part or in whole, for any purpose and by any means, provided that credit is given.
Date: 2014.02.01. Permanent ID of this document: 5e940efc30c7dc872154fb494be7b5fd.}\\
Email/XMPP: martin@mchomenet.com}
\date{}
\begin{document}
\maketitle
\singlespacing

\begin{abstract}
There is a multitude of licenses out there for a software developer to choose from, but a lot of programmers would prefer to not have to have a legal degree in order to understand them and would rather just have their code out in the public being used. This paper examines a few of the more popular and common software licenses available and compares their conditions from the perspective that a developer wanted to scratch their proverbial itch, have their code be used and maybe even get credit for it. This paper attempts to clarify which license a developer would benefit from the most if they just want to have their code used, have their name out there and not be sued if anything breaks.
\end{abstract}

\noindent{\bf Index Terms}: open source, licenses, software, freedom, restriction, comparison.


\section{Introduction}
The history of Open Source is a long story of ideals and politics and involves developers being creative and program sources being useful. Before proprietary software became common place, it was not unusual for developers to release their code for other developers to use without a second thought as to how they might use it, so long as they found it useful and may even help save the developer some time in a project.

Software was shared via floppy disk at computer meet and greets, via the early internet and everything in between. Developers did not need to think about or even worry about software licenses, everything was implicitly in the public domain, without it necessarily being dedicated to the Public Domain, due to a lack of relevant licenses and developer apathy towards enforcement\cite{earlyopensource}. Code was shared, original authors names acknowledged and programs were built through the indirect contributions of others. Then proprietary software came along and complicated matters.

Proprietary software was about closing up your sources so others could not freely learn from what you had done and was about not sharing the results of software discoveries. But there still remained developers who wanted to contribute to the common pool of useful software and allow others to share in the discovery again of algorithms and interesting ideas.

To a developer wishing to get his or her code in the eyes of fellow developers, choosing an open source license can seem daunting and overwhelming. Not too many developers have training in legal matters and many of them just want to code, either for themselves to later release, or perhaps to contribute to an existing project. This leaves the developer in a dilemma for which license to apply to their work.

Placed in these circumstances, the developer who just wants to get their code and their name out in the public might just release their code under a standard "By Firstname Lastname" or "Copyright Firstname Lastname" and then presume this is enough for others to use. However, this is a false presumption and may not mean what the developer thinks it means.

Under default copyright law, a simple "Copyright Firstname Lastname" already implies "All rights reserved"\cite{berneauto}, and the receiver can only read your code and not do anything with it at all without your permission. This is why a license, even a simple one, is needed so that the receiver will know what they can do.

A long and tedious license is not required either, otherwise there might end up being more lines in the license then there are of actual code and therefore, a simpler one might suffice.

A lot of developers just want to get their name out there, have people use the code, maybe send back a kudos and perhaps some tweaks, and to not get into trouble if anyone does anything silly with the code. They want fellow developers to feel free to use their code however they want if they find it useful. 

Fortunately there is a way to do this without long and tedious licenses, without needing legal training and in a simple way that the developer can release their code, not have to worry about it and get back to coding. This paper examines the various ways a computer programmer can go about this by looking at the several options available, in order to find the simplest and most practical method to achieve this.

\section{Evaluations}
How might one find that delicate balance in a software license that brings with it simplicity, freedom for the developer, freedom for the user and is most practical. Here we look at the GNU General Public License v2, the GNU General Public License v3, the MIT license and the 2-clause BSD license.

The ISC license, which is virtually identical to the MIT license, is included in the MIT license examination while the Apache v2 license is not discussed here as it delves into patents, which is for another paper. Here we only delve into expression of an idea as shown in the form of code that developers use to produce software.

\subsection{Summary of the Licenses}
\subsubsection{GPL v2}
The GNU General Public License Version 2 comes from the mind of Richard Stallman of the Free Software Foundation. Released in June 1991 By Stallman, it has the following:

\textbf{Advantages}
\begin{itemize}
\item Allows commercial use
\item You are free to modify the source
\item You are free to re-distribute the program, both modified and unmodified.
\item You can place a warranty on the software.
\end{itemize}
\textbf{Disadvantages} are:
\begin{itemize}
\item You must include a copy of the original sources, or provide a written offer for the user to be able to get the sources from you.
\item You must also provide any modified copies of the source code you used to build your program.
\end{itemize}
\textbf{Restrictions} include:
\begin{itemize}
\item You can not sub-license the code under any other further restricted license.
\item You can not sue the author(s) for damages.
\item You must provide a copy of the license with all distributions.
\end{itemize}
This license ensures the maximum freedom of the code to remain open (viewable) and free (as in speech), by placing restrictions upon users and their usage of the code.

\subsubsection{GPL v3}
The GNU General Public License Version 3 is a sequel to Version 2 previously discussed, and comes from a list of refinements that Richard Stallman felt would better address new technological changes and improvements since Version 2 was released. Written by Richard Stallman with assistance from Eben Moglen\cite{gpl3} and published on 29th June 2007\cite{gpl3published}, it features the following:

\textbf{Advantages}
\begin{itemize}
\item Allows commercial use
\item You are free to modify the source
\item You are free to re-distribute the program, both modified and unmodified.
\item You can place a warranty on the software.
\end{itemize}
\textbf{Disadvantages}
\begin{itemize}
\item You must include a copy of the original sources, or provide a written offer for the user to be able to get the sources from you.
\item You must also provide any modified copies of the source code you used to build your program.
\item You must explicitly state your changes made to the source code.
\end{itemize}
\textbf{Restrictions}
\begin{itemize}
\item You can not sub-license the code under any other further restricted license.
\item You can not sue the author(s) for damages.
\item You must provide a copy of the license with all distributions.
\item If you use this software in an embedded system which requires security keys verified in order to operate and work, you must provide the source to the verifier and/or key generator. You can not use this software with hardware Digital Rights Management, for example.
\end{itemize}
This license ensures the maximum freedom of the code to remain open (viewable) and free (as in speech), by placing restrictions upon users and their usage of the code, be it normal software, secured firmware and/or embedded.

\subsubsection{2-clause BSD}
The Simplified aka 2-Clause BSD license, created by the University at Berkeley and named after their operating system which was called the Berkeley Software Distribution (of Unix) is a basic license, practically identical in spirit and intent as the MIT license. It features:

\textbf{Advantages}
\begin{itemize}
\item Allows commercial use
\item You are free to modify the source
\item You are free to distribute the program, whether modified or not.
\item You can sub-license the code under any other license you choose, provided they have similar requirements in attribution.
\item You can provide a warranty on the end product.
\end{itemize}
\textbf{Disadvantages}
\begin{itemize}
\item You must include a copy of the license which acknowledges the original author(s) and has the disclaimer.
\end{itemize}
\textbf{Restrictions}
\begin{itemize}
\item You can not sue the author(s) for damages.
\end{itemize}

\subsubsection{MIT}
The MIT license, also famous for being associated with the X Window System (also created at MIT), the MIT license is a very basic license that is very easy to understand and follow. Written by MIT, features include:

\textbf{Advantages}:
\begin{itemize}
\item Allows commercial use
\item You are free to modify the source
\item You are free to distribute the program, whether modified or not.
\item You can sub-license the code under any other license you choose, provided they have similar requirements in attribution.
\end{itemize}
\textbf{Disadvantages}
\begin{itemize}
\item You must include a copy of the license which acknowledges the original author(s) and has the disclaimer.
\end{itemize}
\textbf{Restrictions}
\begin{itemize}
\item You can not sue the author(s) for damages.
\end{itemize}
The MIT license is functionally equivalent to the 2-clause BSD license, except with wording removed that was made redundant with the Berne Convention. Another license, the ISC license, is functionally identical to the MIT license as well.

\subsubsection{Others}
There are other "licenses" which are available which offer more and more freedoms, both for the developer, the user and the code. They include:
\begin{itemize}
\item Unlicense - Essentially Public Domain but with an implicit fall-back to an MIT-like license if the user is in a jurisdiction that does not recognise the voluntary abandonment of an author's rights.
\item Public Domain - This is the ultimate permission for source code: Do what you want, you already have every permission there is. You already own it. Does not apply to all jurisdictions and is said to be "tricky". Not applicable in Germany and most parts of Europe.
\item Basic Software License - A very simple license. Contains a copyright line, a statement saying "Do with this as you wish", and a disclaimer against liability. For those who are looking for the next best thing to public domain, but something shorter than the Unlicense.
\end{itemize}

\subsection{Evaluating the Philosophies}
\subsubsection{Source to Study}
At a minimum, one might find just having the right to be able to read the source to study it and ensure there is nothing malicious in it might be acceptable. Following on from that, a user would then rely upon copyright law to allow them to make their desired changes to the source for personal use. If the developer wished to share those changes with others, it would be necessary to make a list of these changes so others can re-produce them, or provide them in a "patch" file that other developers can apply to the source to re-create those changes automatically. This is the method Daniel J. Bernstein originally followed with his qmail software suite \cite{djb}.

\subsubsection{Source to Modify and Share}
One step beyond the above is to allow fellow developers to share their changes freely with others, without any restriction. If they add a new feature, modify an existing feature or just fix a bug, it is important that developers feel they can freely send those changes back to the original developer, or just re-distribute their modified version of the code. Either way, the improved program gets released and contributes to a universal pool of useful software.

\subsubsection{Free as in Price}
Capitalism states\cite{supplydemand} that where there is a demand, there is a market, and where there is a market, there is always a certain high price that people are willing to pay, and sellers should always try to charge the most they can get away with and that the market can bare.

This does not help developing and third-world countries, students and those on very low income who are just trying to get ahead and make their way in the world. So many people find it very useful to be able to just purchase computer hardware, and then to load it with free as in price software. If they want to learn how the software works then it helps if the software source code is open. In some cases, this works well for those who can not afford sometimes expensive guides and books to learn from.

Free as in price means everyone can get the software, either completely free or for just the cost of distribution.

\subsubsection{Free as in Freedom}
Like free speech, developers and users alike appreciate that when they receive a program, whether as source and binary or just source, that it allows them to be able to modify it to suit their needs, fix any bugs, add any new features and then to be able to share those changes with others. This is in stark contrast with proprietary software which is provided in binary form only, and you have to rely upon the original developer(s) to fix bugs and add new features.

\subsubsection{Freedom of...}
Freedom of the Source, Freedom of the Developer and Freedom of the User is where the ideals behind open source licenses really diverge. Freedom of the code is best emphasised in the GPL2 and GPL3 licenses, Freedom of the User is a mix between GPL2, GPL3 and MIT/BSD and Freedom of the Developer is best met with the BSD/MIT licenses.

Guaranteeing that the source code to a program can be studied, modified and that those changes will also be kept open so that others can use them as well is the basic concept behind the GNU General Public License. By forcing the developer to keep their changes public as well, this ensures that the source of the program, regardless of who has it and what they may or may not have done to it, is always available for people to use. Though an admirable intent, it does means that if a developer extends the software with some trade secrets, those secrets must now be known and revealed to whomever the software is distributed to.

Any software linked to that program must also have its source code open with the same restrictions. Sometimes, people don't make any changes at all beyond compiling it and making a larger system combined with other pieces to make a more complete environment. However, even if no changes are made at all, the distributor is still required to host the source code themselves. This leads to 1,000's of copies of the same software being hosted and can be considered redundant, especially if the original developer has reliable hosting for the code.

This is in contrast to the simpler MIT and BSD licenses which state that the receiving developer can use the code, modify it, re-write it, do whatever they like with it provided that the original developers name is left in tact within the code as acknowledgement of original authorship and that no-one tries to sue them. This provides a much easier way for anyone, individual or corporation, to be able to benefit from the code. Many developers find programming to be a fun and intellectual exercise which sets out to achieve a purpose, after which, they might release and would get a sense of satisfaction knowing that they have helped other developers have to do less in order to make a great product or system.

\section{Results and Discussion}
Can one gain a freedom from a restriction? Can having more restrictions on a license make it seem more free? Code is not a living being, so it has no rights granted to it. So it comes down to what the developer permits you to do with it. The MIT license is clearly a license, whereas several parts of the GPL make it seem more like a contract; a contract that is enacted upon your receiving it.

This contract states what you can and can't do with it, what you must do when making changes and what you are required to do if sharing it or re-distributing it to others. It does this via a long list of detailed conditions that many feel you need a law degree just to read and comprehend.

Whereas, the MIT license is a license which says to the receiving developer: use this code however you want, just leave my name in the code, do not sue me if something in here breaks and make sure a copy of this list of permissions and do-not-sue-me notice stays with the software.

Take the above into account, from the perspective of freedom to the user and developer to be able use and benefit from the code and it's capabilities, and it would appear that the MIT and BSD license (with the similarly worded ISC license) allow the most freedom to use and enjoy the code in whatever way the user sees fit.

No-one likes being forced to do something they do not particular want to do, and therefore anything they are forced to do may have a lower quality output from obligation. However, if giving back was optional and keeping in the spirit of how a developer meant the code to be used and shared, they are more likely to get less submissions back, but they would be of higher quality as there would most likely be more pride in the code and it's functionality. This is in contrast to the forced giving back of the GPL which would get a much higher level of submissions, but many of them would be of a lower quality and may only be of benefit to the developer who implemented those changes to suit their own environment and would be of little practical use to anyone else.

\section{Conclusion}
The open source concept is about making things work, being proud of your work, scratching the developers' proverbial itch and getting things done as efficiently as possible while getting things done as fast as possible. The MIT and BSD license allows you to be proud of your work and allows you to help others with higher quality feedback.

The GPL and the goals behind the Free Software Foundation brought back the idea to people that sharing your source was an alright thing to do and it helped re-introduce the idea of bringing a community around a project to support it and give back would make it bigger and better. In this it succeeded.

The next step is to remove the restrictions which force freedom of the code, and allow people to use the code how they want and to not require them to give back, they should give back only if they want to. This would likely lead to a higher quality of submissions that would actually be used.

If more and more developers used the 2-clause BSD license for their next library or project, it would contribute to a universal pool of useful software that fellow developers could build upon and use. In turn, they would also release their projects similarly, providing other developers again a resource of useful software.

Freedoms born of a restriction may not necessarily be the kind of freedoms that a developer is aiming for, but a freedom from restrictions will allow developers to be more productive, both from the donor's point of view as well as the receiver. Freedom to use a piece of useful code without needing to get caught up in distracting matters with the minimum of fuss is the kind of freedom that future developers are sure to appreciate.

\bibliographystyle{plain}

\section*{Acknowledgements}
Special thanks to Kerry-Anne COLEMAN for her support and encouragement while I put this material together and to Peter GREEN for proofreading assistance and suggestions.

\section*{Appendix}
Here is a template 2-clause BSD license.

---------
\singlespacing
Copyright (c) <YEAR>, <OWNER>
All rights reserved.

Redistribution and use in source and binary forms, with or without modification, are permitted provided that the following conditions are met:

1. Redistributions of source code must retain the above copyright notice, this list of conditions and the following disclaimer.

2. Redistributions in binary form must reproduce the above copyright notice, this list of conditions and the following disclaimer in the documentation and/or other materials provided with the distribution.

THIS SOFTWARE IS PROVIDED BY THE COPYRIGHT HOLDERS AND CONTRIBUTORS "AS IS" AND ANY EXPRESS OR IMPLIED WARRANTIES, INCLUDING, BUT NOT LIMITED TO, THE IMPLIED WARRANTIES OF MERCHANTABILITY AND FITNESS FOR A PARTICULAR PURPOSE ARE DISCLAIMED. IN NO EVENT SHALL THE COPYRIGHT HOLDER OR CONTRIBUTORS BE LIABLE FOR ANY DIRECT, INDIRECT, INCIDENTAL, SPECIAL, EXEMPLARY, OR CONSEQUENTIAL DAMAGES (INCLUDING, BUT NOT LIMITED TO, PROCUREMENT OF SUBSTITUTE GOODS OR SERVICES; LOSS OF USE, DATA, OR PROFITS; OR BUSINESS INTERRUPTION) HOWEVER CAUSED AND ON ANY THEORY OF LIABILITY, WHETHER IN CONTRACT, STRICT LIABILITY, OR TORT (INCLUDING NEGLIGENCE OR OTHERWISE) ARISING IN ANY WAY OUT OF THE USE OF THIS SOFTWARE, EVEN IF ADVISED OF THE POSSIBILITY OF SUCH DAMAGE.


\end{document}